\documentclass[fleqn,usenatbib]{mnras}

\usepackage{newtxtext,newtxmath}

\usepackage[T1]{fontenc}
\usepackage{ae,aecompl}

\usepackage{graphicx}	
\usepackage{amsmath}	
\usepackage{amssymb}	

\title[A possible UFO in Swift J0243.6+6124]{\textit{Chandra} reveals a possible ultra-fast outflow in the\\ super-Eddington Be/X-ray binary Swift J0243.6+6124}

\author[J. van den Eijnden et al.]{
J. van den Eijnden$^{1}$\thanks{E-mail: a.j.vandeneijnden@uva.nl (JvdE)},
N. Degenaar$^{1}$, N. S. Schulz$^{2}$, M. A. Nowak$^{3}$, R. Wijnands$^{1}$, \newauthor T. D. Russell$^{1}$, J. V. Hern\'andez Santisteban$^{1}$, A. Bahramian$^{4}$, T. J. Maccarone$^{5}$, \newauthor J. A. Kennea$^{6}$, C. O. Heinke$^{7}$
\\
$^{1}$Anton Pannekoek Institute for Astronomy, University of Amsterdam, Science Park 904, 1098 XH Amsterdam, The Netherlands\\
$^{2}$Kavli Institute for Astrophysics \& Space Research, MIT, 70 Vassar Street, Cambridge, MA 02139, USA\\
$^{3}$Physics Department, CB 1105, Washington University, One Brookings Drive, St. Louis, MO 63130-4899, USA\\
$^{4}$International Centre for Radio Astronomy Research -- Curtin University, GPO Box U1987, Perth, WA 6845, Australia\\
$^{5}$Physics Department, Texas Tech University, PO Box 41051, Lubbock, TX 79409, USA\\
$^{6}$Department of Astronomy and Astrophysics, The Pennsylvania State University, University Park, PA 16802, USA\\
$^{7}$Department of Physics, University of Alberta, CCIS 4-181, Edmonton, AB T6G 2E1, Canada
}

\date{Accepted XXX. Received YYY; in original form ZZZ}

\pubyear{2018}

\begin{document}
\label{firstpage}
\pagerange{\pageref{firstpage}--\pageref{lastpage}}
\maketitle

\begin{abstract}
Accretion at super-Eddington rates is expected to be accompanied by strong outflows. Such outflows are observed in Galactic X-ray binaries and extragalactic Ultra-luminous X-ray sources (ULXs). However, due to their large source distances, ULX outflows are challenging to detect and study in detail. Galactic neutron stars accreting from a Be-star companion at super-Eddington rates show many similarities to ULX pulsars, and therefore offer an alternative approach to study outflows in this accretion regime. Here, we present \textit{Chandra} high-resolution spectroscopy of such a super-Eddington accreting neutron star, Swift J0243.6+6124, to search for wind outflow signatures during the peak of its 2017/2018 giant outburst. We detect narrow emission features at rest from Ne, Mg, S, Si, and Fe. In addition, we detect a collection of absorption features which can be identified in two ways: either as all Fe transitions at rest (with a possible contribution from Mg), or a combination of three blue-shifted Ne and Mg lines at $\sim 0.22c$, while the remaining lines are at rest. The second scenario would imply an outflow with a velocity similar to those seen in ULXs, including the ULX pulsar NGC 300 ULX-1. This result would also imply that Swift J0243.6+6124 launches both a jet, detected in radio and reported previously, and an ultra-fast wind outflow simultaneously at super-Eddington accretion rates. 

\end{abstract}

\begin{keywords}
accretion, accretion discs -- stars: neutron -- X-rays: binaries -- pulsars: individual: Swift J0243.6+6124
\end{keywords}



\section{Introduction}

The accretion and subsequent ejection of matter is a ubiquitous process in the Universe, occurring in objects ranging from young stellar objects and planet-forming systems \citep[e.g.][]{kuiper15,beltran15} to X-ray binaries \citep[see e.g.][for overviews]{fender04,migliari06} and active galactic nuclei \citep[e.g.][]{merloni03,falcke04}. In X-ray binary systems (XRBs) a stellar-mass compact object, either a black hole or a neutron star, accretes from an companion star in a close orbit.  X-ray binaries come in numerous classes, with different combinations of compact object type, and mass and type of companion. Additionally, the accretion can take place through various channels \citep{frank92}, triggered by for instance Roche-lobe overflow of the donor \citep{kuiper41,paczynski71}, a stellar wind \citep[e.g.][]{reig11} or the movement of the compact object through the circumstellar disk of the donor star \citep{okazaki01}. 

Accretion in X-ray binaries is often accompanied by the ejection of matter, either through disk winds or via jets. The latter are strongly-collimated outflows traveling near the speed of light, launched from the inner accretion flow, while winds are launched further out from the accreting object at lower velocities (ranging from hundreds of km/s to $\sim 0.3c$) and with wider opening angles. In X-ray binaries, winds can carry away a large fraction of the mass from the accretion flow \citep{neilsen09,ponti12}, possibly triggering instabilities in the flow \citep{begelman83,munozdarias16} and potentially affecting the outburst profiles of transient sources \citep{btetarenko18}. Similarly, jets can remove accretion power from the X-ray binary and deposit large amounts of energy in the surrounding interstellar medium \citep{fabrika04,fender05,gallo05,pakull10}.

In low-mass X-ray binaries (LMXBs) -- XRBs with a low-mass (i.e. $\lesssim 1 M_{\odot}$) donor -- accreting below the Eddington limit, observational and theoretical work suggests that compact jets and disk winds are generally not launched simultaneously \citep{miller06,neilsen09,ponti12,higginbottom15,bianchi17}. Steady jets are typically seen at relatively low X-ray luminosities \citep{fender04,fender09} during hard X-ray spectral states \citep[see e.g][for an overview of spectral states]{gilfanov10}. During the soft state, jets in black hole LMXBs appear to be quenched \citep[e.g.][]{fender09, coriat11}, while the picture is more complicated for neutron star LMXBs \citep[e.g.][]{migliari04,millerjones10,migliari11c,fender16,gusinskaia17}. Winds, on the contrary, are typically not seen in the hard state \citep{ponti12,neilsen13}, although see \citet{xu18} and \citet{maccarone16} for possible counterexamples. 

Outflows can alternatively be studied in X-ray binaries accreting around or above the Eddington luminosity; in such sources, strong outflows are expected due to the enhanced radiation pressure exerted on the accretion flow \citep{shakura73,ohsuga11,mckinney14,mckinney15,hashizume15,kingmuldrew16}. Famous examples of X-ray binaries launching strong outflows in this regime are the black-hole LMXBs GRS 1915+105 \citep{mirabel94,neilsen09} and V404 Cygni \citep{munozdarias16,tetarenko17_v404jets}, and the accreting neutron star Cir X-1 \citep{brandt00}. During super-Eddington X-ray binary states, the apparent dichotomy between winds and jets can also break down; for instance, black holes in XRBs and Z-sources -- a subset of neutron star LMXBs categorized based on their X-ray color-color diagram \citep{hasinger89} and accreting around the Eddington limit -- are thought to launch a wind and jet simultaneously at such accretion rates \citep{homan16,allen18}.

In this regard, Ultra-luminous X-ray sources (ULXs) are particularly interesting \citep[see][for a recent review]{kaaret17}. These extragalactic X-ray emitters have X-ray luminosities (greatly) exceeding $\sim 10^{39}$ erg/s, or the Eddington luminosity of a $10 M_{\odot}$ black hole. Recently, a handful of ULXs has been identified as accreting neutron stars through the detection of pulsations \citep{bachetti14,furst16b,israel17a,israel17b,carpano18}, with several additional candidates found through possible cyclotron resonance scattering features \citep{brightman18,koliopanos19,walton18b}. This confirms the super-Eddington nature of at least a fraction of ULXs. While it is unclear what fraction of ULXs contains a pulsar, both theoretical \citep{king01b,king16b} and observational studies \citep{koliopanos17,walton18c} suggest it could be substantial. 

Theoretically, outflows have often been suggested to explain the soft spectra of (some) ULXs \citep{king01,begelman02,king03,gladstone09,feng11,urquhart16}. From the observational side, jets have been observed indirectly from these sources through their impact on the surrounding medium \citep{middleton13,cseh14,cseh15a,cseh15b,mezcua15} and inferred from radio detections \citep[e.g.][]{kaaret03,webb12,mezcua13}. Winds have been seen through X-ray \citep{pinto16} and optical \citep{zepf08} spectroscopy of several targets, including one ULX pulsar \citep[NGC 300 ULX-1;][]{kosec18b}. However, the extragalactic nature of ULXs complicates the study of their outflows. Scaling for instance typical radio luminosities of compact jets launched by black holes accreting around the Eddington luminosity \citep[e.g.][]{gallo18} to Mpc distances, yields flux densities at best around the detection limit for current generation radio arrays. Similarly, the detection of winds through X-ray spectroscopy is limited by the low number of counts in the X-ray grating spectra. As a result, most high-resolution X-ray spectra of ULXs currently available in the archive are not sensitive enough to reveal any wind signatures \citep{kosec18a}.

With their smaller distances, X-ray binaries in the Milky Way or the Small Magellanic Cloud may offer a valuable alternative avenue to study these super-Eddington accretion states at higher signal-to-noise ratio (although the sample of such sources is limited by the smaller volume and extreme count rates can introduce calibration issues, as discussed later). In particular, neutron star Be/X-ray binaries, wherein the donor is a Be-star, show many similarities to the known ULX pulsars \citep{mushtukov15,koliopanos17}: strong ($\geq 10^{12}$ G) magnetic fields and slow spins (i.e. periods on the order of seconds). Importantly, Be/X-ray binaries can also show super-Eddington accretion rates during the peaks of their giant outbursts \citep[e.g][]{reig11}. 

In September 2017, the \textit{Swift} satellite discovered the new neutron star Be/X-ray binary Swift J0243.6+6124 \citep[hereafter Sw J0243;]{kennea17}: a strongly-magnetized neutron star \citep[e.g.~B~$>~10^{12}$~G;]{tsygankov18} with a $\sim 9.8$ second spin period \citep{kennea17}. It reached super-Eddington X-ray luminosities during the peak of its outburst \citep{vandeneijnden18d} and has been referred to as the first Galactic ULX pulsar by both \citet{tsygankov18} and \citet{wilson18}. Through \textit{Very Large Array (VLA)} radio and \textit{Swift} X-ray monitoring, Sw J0243 was found to launch a relativistic jet during its super-Eddington state \citep{vandeneijnden18d}, as well as at lower accreting rates \citep{vandeneijnden19}. This jet detection constituted the first from a strongly-magnetized neutron star, contrary to the predictions of neutron star jet formation theory \citep{blandford82,massi08}.    

Here, we present \textit{Chandra} high-resolution gratings X-ray spectroscopy of Sw J0243 in its super-Eddington state. We find evidence for a wind with a velocity of $\sim 0.22$c through the detection of blue-shifted absorption features (Section \ref{sec:ident_ab2}), similar to those detected in ULXs \citep{kosec18b}. If these features indeed arise from a wind, this would imply both a jet and a wind are launched simultaneously by Sw J0243 during its super-Eddington state. 

\section{Observations and Data Reduction}
\label{sec:obs}

\begin{figure*}
	\includegraphics[width=\textwidth]{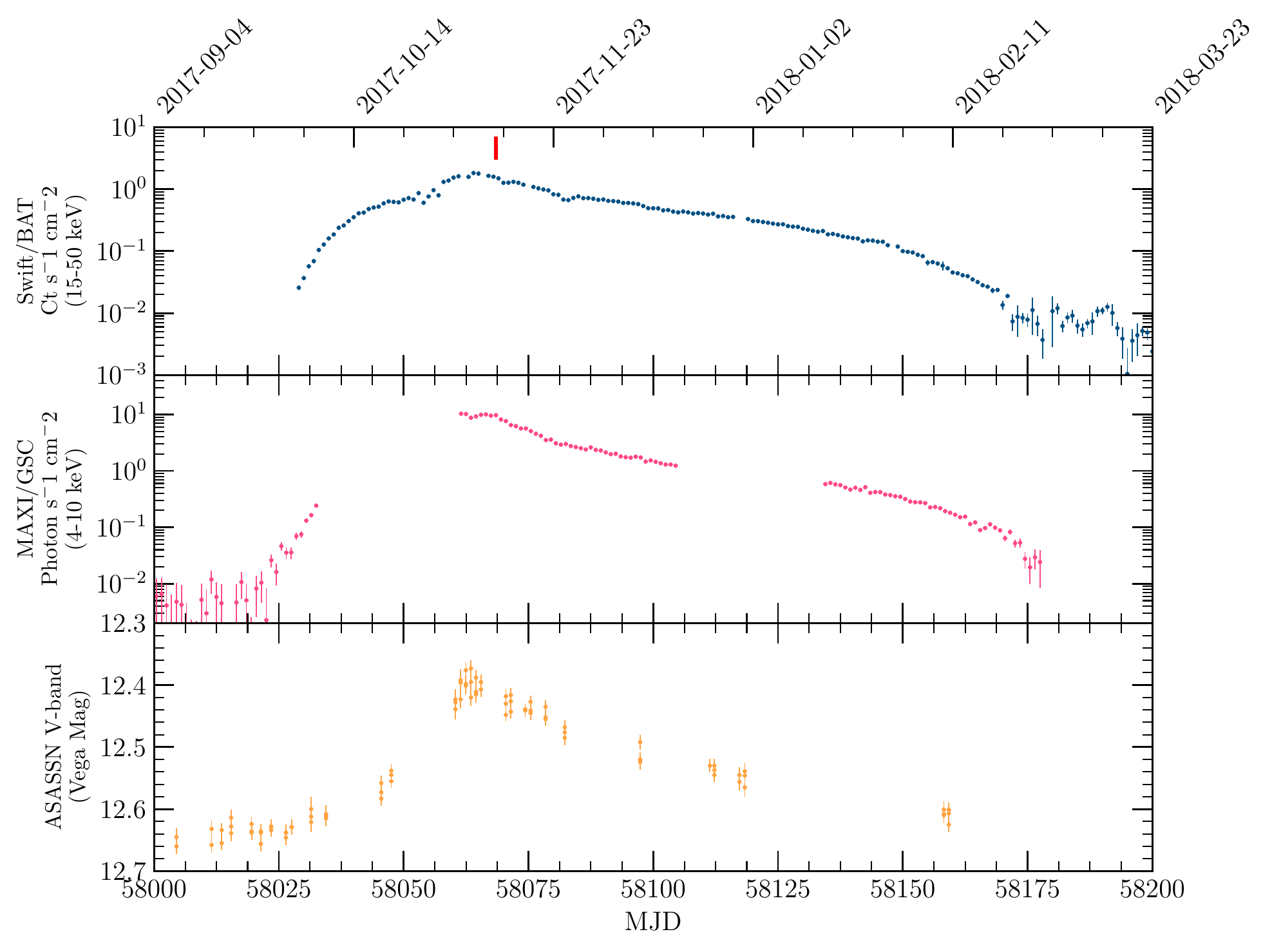}
    \caption{Multi-band light curves of the 2017/2018 outburst of Sw J0243. The top panel shows the \textit{Swift}/BAT $15$--$50$ keV count rate, the middle panel shows the \textit{MAXI}/GSC $4$--$10$ keV count rate, and the bottom panel shows the ASAS-SN V-band magnitude (Vega Mag). The Chandra epoch is indicated with the red line in the top panel.}
    \label{fig:lightcurve}
\end{figure*}

In Figure \ref{fig:lightcurve}, we show the X-ray and optical light curves of the 2017/2018 outburst of Sw J0243. \textit{Chandra} observed the target around the peak of this giant outburst. While the distance to Sw J0243 is not precisely known, the Gaia DR2 implies a minimum of $5$ kpc at $99\%$ confidence \citep{vandeneijnden18d}. Given this minimum distance, the X-ray luminosity during the stage of the outburst around the \textit{Chandra} epoch exceeded $10^{39}$ erg/s in the $0.5-10$ keV band \citep{wilson18,vandeneijnden18d}. As the theoretical Eddington luminosity for a neutron star is $2\times10^{38}$ erg/s, this luminosity implies a firmly super-Eddington accreting rate. 

\emph{Chandra} performed Director's Discretionary Observations of Sw J0243 on 11 November 2017 (MJD 58068) for $\sim$ 25 ks of exposure (ObsID 20859; PI Degenaar) with the high energy transmission grating spectrometer (HETGS). Given the extreme flux of the source, the observation setup had to mitigate photon pile-up as well as minimize telemetry saturation. The telescope aimpoint was moved to the CCD readout where the zero order dithered between the framestore and a few CCD rows and thus only partially covered the active CCD area. In this configuration only two grating dispersion arms are recorded, the medium energy grating (MEG) +1st order and the high energy grating (HEG) -1st order plus their higher order dispersions. The observation was recorded in continuous clocking mode (CC-mode) with a fast readout time of 3.85 msec to effectively mitigate photon pile-up in the dispersed spectra. The data were transmitted via GRADED mode which includes onboard event grading and grade summing. This results in some loss of data information and aspects of calibration become more approximate. While this mostly preserves the detection of discrete line features and edges, it does affect the calibration of the spectral continua in the first order spectra see also \citealt{schulz09} for another example of CC-mode observations where mainly the discrete line features are preserved, and \citealt{miller03} for the analysis of CC-mode spectra of the accreting black hole XTE J1550-564). 

The observation data were re-processed via CIAO 4.9 using the latest calibration product at the time (CALDBv4.7.7). Changes in more recent versions did not have any impact on the analysis at the time of submission. We used the \emph{run\_pipe} thread within the \emph{tgcat} package in \emph{ISIS}\footnote{see \url{http://space.mit.edu/ASC/ISIS}}. Under normal circumstances this determines the wavelength scale to the accuracy of a quarter of a resolution element, i.e. 0.005 \AA\ for MEG and 0.002 \AA\ for HEG. However due to the fact that the zero order is piled as well as that it dithers on and off the chip likely adds another quarter in systematic uncertainty. In gratings the dispersion scale is linear in wavelength and all line analysis will be done in wavelength space. We used standard wavelength redistribution matrices (RMFs) and generated ancillerary response files (ARFs) applying 
provided aspect solutions, bad pixel maps, and CCD window filters\footnote{The extracted HEG and MEG spectra and responses are available from the author upon request.}.

Given the possible continuum issues due to the extreme count rates, we also extracted two \textit{Swift} XRT spectra to compare with the \textit{Chandra} HEG and MEG spectra. We used the Swift XRT data products generator \citep{evans07}\footnote{\href{http://www.swift.ac.uk/user\_objects/}{http://www.swift.ac.uk/user\_objects/}} to extract the \textit{Swift} spectra taken on 10 and 13 November 2017 (MJDs 58067 and 58070, with ObsIDs 10336023 and 10336024, respectively). Both observations were taken in WT-mode, which can deal with high count rates, and the data products generator automatically corrects for any pile-up issues for very bright sources. 

\section{Analysis \& Methods}

\subsection{Continuum analysis and spline modeling}
\label{sec:continuum}
\begin{figure}
	\includegraphics[width=\columnwidth]{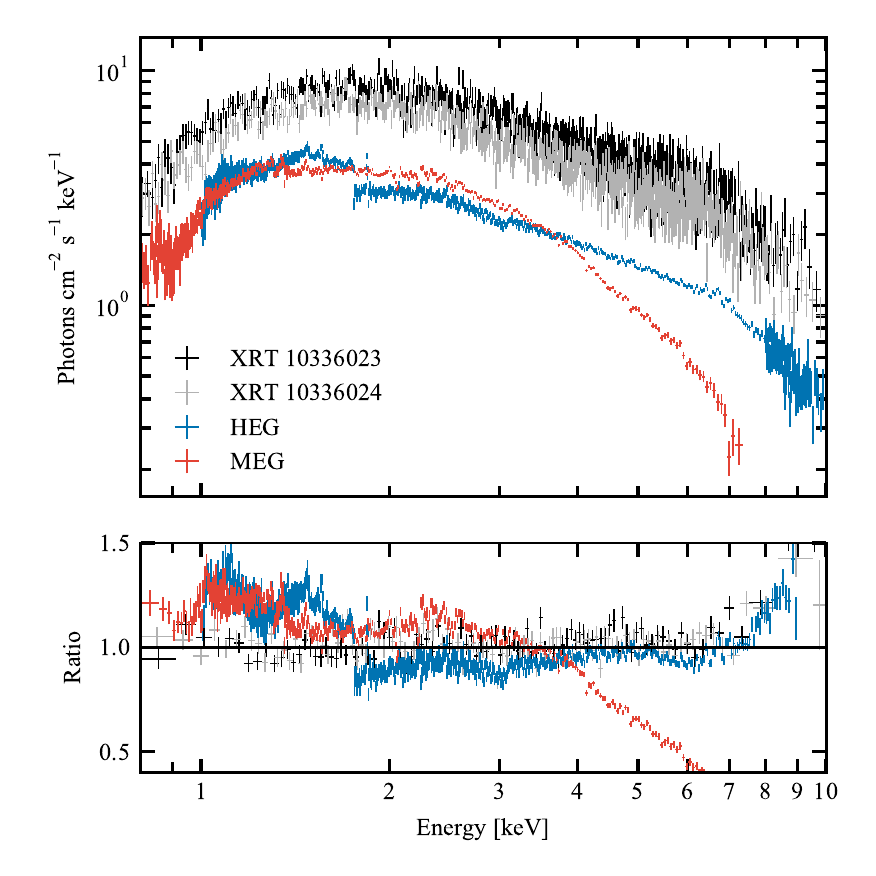}
    \caption{Unfolded \textit{Chandra} HEG (blue) and MEG (red) spectra, together with the \textit{Swift} XRT spectrum of the preceding (1 day prior) and following (2 days later) epochs. While the two \textit{Swift} spectra are similarly shaped, the MEG and HEG spectra show large deviations both from each other and from the XRT spectra. This can also be seen in bottom panel, which shows the data-to-model ratio for a \textsc{tbabs*(bbodyrad+po)} model fit jointly to the XRT spectra. }
    \label{fig:crappycontinuum}
\end{figure}

In Figure \ref{fig:crappycontinuum} (top), we show both the \textit{Chandra} MEG and HEG spectra with the two \textit{Swift} XRT spectra taken before and after the \textit{Chandra} epoch. As described in Section \ref{sec:obs}, the unconventional HETGS observing setup required by the extreme flux can affect the calibration of the continuum in the first order spectra. This is clearly at play in our observations of Sw J0243: inaccuracies in the HETGS continua are obvious from the disparity between the MEG and HEG detectors, the large jump in the HEG spectrum around $\sim 7$~\AA~($\sim 1.77$ keV), and both the offset and difference in spectral shape between the \textit{Swift} and \textit{Chandra} spectra. The latter is highlighted by the bottom panel of Figure \ref{fig:crappycontinuum}, where we show the ratio of the MEG and HEG data to a simple \textsc{tbabs*(bbodyrad+po)} model fit jointly to the XRT spectra with \textsc{xspec} \citep{arnaud96}. 

However, while the calibration of the continuum shape is affected by the observational setup, discrete line features and edges are preserved \citep{schulz09}. Indeed, individual narrow features remain. In the HEG spectrum, for instance, a clear Fe K complex around $\sim 6.4$--$7$ keV is visible, as discussed in detail in Section \ref{sec:ident_em}. Therefore, one can still search for narrow emission and absorption features from, for instance, outflows or donor star material. As the offset between the \textit{Chandra} and \textit{Swift} spectra demonstrates, the HEG and MEG data will not provide an accurate measurement of the flux. As a result, conventional line strength measures such as the equivalent width or normalization of any such narrow features will not be accurate. Furthermore, without an accurate continuum measurement, physically motivated modelling is challenging for any model that contains a (pseudo)-continuum component. But despite these restrictions, the presence of narrow emission and absorption features can still be tested and give valuable physical information about the system. 

Any search algorithm for individual narrow X-ray spectral lines requires an accurate description of the continuum. This holds especially for the approach that we adopt for Sw J0243, which was originally developed for the detection of faint features in ULX spectra by \citet{pinto16} and is introduced in Section \ref{sec:linesearch} \citep[for an illustration of the effects of a poorly modeled continuum on the detection and significances of narrow line features, see also][]{vandeneijnden18c}. However, this line search approach only requires an accurate shape of the continuum model to find narrow deviations from; this continuum model does not necessarily have to be physically motivated. A example of this can be found in \citet{grinberg17}, where a \textit{Chandra} spectrum of the HMXB Vela X-1 is modeled; in that work, the continuum consists of four independent power law models, which are not physically motivated, fitted over a limited wavelength range. However, these models do provide an accurate description of the underlying continuum shape and allow for the search and identification of narrow line features. A similar mathematical approach to model the continuum can be found in \citet{yao09}, where a combination of broad Gaussians makes up the continuum model. Therefore, instead of fitting the \textit{Chandra} Sw J0243 continuum with physical models -- which is not possible for the full spectral range, as shown in Figure \ref{fig:crappycontinuum} -- we apply a spline interpolation as the continuum instead. 

To calculate the spline interpolations of the HEG and MEG spectra, we first used \textsc{xspec} to write out the flux as a function of wavelength. We then choose the step size of the interpolation -- as we aim to search for deviations from the spline continuum, we should not interpolate every spectral bin but instead bins separated by a fixed wavelength -- and defined a wavelength grid with such steps on the considered wavelength range (note that, therefore, this grid is \textit{not} the same as wavelength bins of the HEG and MEG detectors). Simply calculating a spline between the fluxes on this grid has the risk of accidentally using either a statistical outlier or a spectral bin inside a narrow line feature as part of the continuum. To prevent this effect, we instead calculated a first continuum estimate with the third degree spline between the fluxes $F_i$ at each grid point in wavelength $\lambda_i$. Then, to obtain the final spline continuum model, we fitted the spline function to the entire spectrum with the $F_i$ values as free parameters, recalculating the spline for each updated combination of $F_i$ and minimizing the $\chi^2$ value between spline and data. Finally, the resulting best-fit continuum spline model was saved as an additive \textsc{xspec fits} table model. 

We calculated separate spline continuum models for the MEG ($2.07$--$13.78$~\AA) and HEG spectra. We used two individual splines for the HEG data, covering $1.55$--$7$~\AA~and $7.1$--$12.4$~\AA, in order to account for the steep jump at $7.06$~\AA. As the presence of a narrow feature in both the HEG and MEG data is important to conclude it is not a mere statistical fluctuation, this implies we exclude the $7$--$7.1$~\AA~range from our entire analysis. We tried different combinations of step sizes ($0.5$, $1$, $1.5$ and $2$~\AA) and spectral binnings (no rebinning, and rebinning to a S/N of $10$ and $50$ per spectral bin) for the calculation of the continuum splines. After a combination of visual inspection and comparison of the continuum $\chi^2$ values, we concluded that a $0.5$~\AA~stepsize and no rebinning provided the most accurate continuum for both the HEG and the MEG data: this combination systematically resulted in the lowest $\chi^2$ values (of the order of $\chi^2_{\nu} \approx 1.2$), while the other combinations (especially with stepsize $\geq 1$ \AA) introduced significant residual structure between the gridpoints interpolated by the spline. 

\subsection{Line search}
\label{sec:linesearch}

We adopt the line search algorithm developed by \citet{pinto16} and refer the reader to that paper and to \citet{kosec18b}, \citet{kosec18a}, and \citet{vandeneijnden18c} for an extensive description of its details. The basic rationale is as follows: after setting a continuum model -- the spline interpolations in the case of Sw J0243 -- we define a grid in wavelength and choose a fixed velocity line width. We then step through the wavelength grid, fitting a single Gaussian function with the fixed velocity width and a free normalization, centered at the grid point. The significance of an emission (i.e. positive normalization) or absorption (negative normalization) line at that wavelength is then recorded as the fitted normalization divided by its one-sigma error. Alternatively, the line significance can also be probed by the improvement in fit statistic (either $\chi^2$ or C-statistic). 

For the correct interpretation of the results of this line search method, several caveats are important to keep in mind. Firstly, despite fitting the spline continuum to cancel the effect of outliers, this continuum does not necessarily describe the entire spectrum accurately. In extreme cases, such as the $7$~\AA~jump in the HEG spectrum, this requires the calculation of multiple splines. However, for less extreme cases, it also implies that care should be taken when considering the physical origin of any suggested lines, and one should carefully inspect the spectra and splines around the possible line features. Furthermore, the returned significances are single-trial estimates, while estimating the number of independent trials is not straightforward. Therefore, line search results of a \textit{single} spectrum should be treated with caution. To reflect this, we do not quote the single-trial significances of any detected lines as actual significances. 

For Sw J0243, we have two simultaneous but independent spectra from two different detectors with different instrument responses. This latter point is important, as any imperfections in the response modeling might appear as deviations from the continuum and resemble a narrow spectral line. However, such features are not expected to appear in both spectra, unless the same response feature is present in both detectors.   

To take these caveats into accounts, we require that any possible spectral lines possess the following properties before considering them as real spectral features of the X-ray binary: (i) the line should be $\gtrsim 3\sigma$ significant (single trial) in both the HEG and the MEG spectrum, (ii) the line should not be located on top of a shared response feature of both detectors, (iii) the continuum model should look accurate around the central wavelength of the line in both detectors and the presence of a spectral feature should hold up to visual inspection of the spectrum, and (iv) the centroid energies, where the significance peaks, of the line in the two detectors should be close: to account for slight statistical deviations between the peak wavelengths and possible small inaccuries in calibration, we require those centroids to be within $\sim 0.01$ keV. Any combination of spectral features adhering to these requirements should of course also fit within a consistent, physically realistic picture of the X-ray binary system and its state. In addition, we also performed a careful visual inspection of the spectra and the line-search results to identify possible features in low S/N parts of the spectra, where lines are less likely to be picked up as significant by the line search.  

\subsection{Robustness of the spline continuum}

As we did not model the continuum with a physical model, but instead with a spline interpolation, we performed several checks of our approach; specifically, we tested whether the line search results, and our inferences, were directly affected by the choice of continuum. For this purpose, we designed two tests: comparing the interpolated continuum with a physical continuum, and slightly varying the step size of the spline grid points.

Figure \ref{fig:crappycontinuum} shows that the \textit{Chandra} and \textit{Swift} spectral shapes do not generally match. However, above $\sim 1.8$ keV, the (blue) HEG spectrum and both \textit{Swift} XRT spectra appear to have a similar shape. Therefore, we fitted two continuum models to these three spectra, using the HEG data between $\sim 1.77$ and $\sim 8$ keV ($7.0$ -- $1.55$~\AA), and both XRT spectra between $1$ and $10$ keV. Using \textsc{xspec}, and assuming abundances from \citet{wilms00} and cross-sections from \citet{verner96}, we fitted both a \textsc{tbabs*po} and a \textsc{tbabs*(bbodyrad+po}) model as simple phenomenological continuum models. In both cases, we included a multiplicative constant to account for offsets between the spectra, while keeping all other parameters tied. Using both these continuum models, we re-applied our line search pipeline with a $500$ and $2000$ km/s line width. This re-analysis finds the same narrow features in the line search, although some residual trends in the line significances remain when using the physical continuum models. These trends suggest that while the HEG and XRT spectra appear similar above $1.8$ keV, small deviations in shape are present. We show these results in more detail in Figure \ref{fig:check_phys} in Appendix \ref{sec:app_checks}. 

Secondly, we re-performed our analysis using a slightly smaller step size for the calculation of the spline continuum -- $0.48$~\AA~instead of $0.5$~\AA -- therefore smoothly connecting different spectral bins with the spline. This check should therefore reveal any imperfections due to the spline by chance connecting statistical outliers and/or narrow lines, instead of probing the continuum. As shown in Figure \ref{fig:check_shift} in Appendix \ref{sec:app_checks} in more detail, the line search results of this test are consistent with our first analysis and do not imply changes in the detected narrow features.  

\section{Results}
\label{sec:results}

\begin{figure*}
	\includegraphics[width=\textwidth]{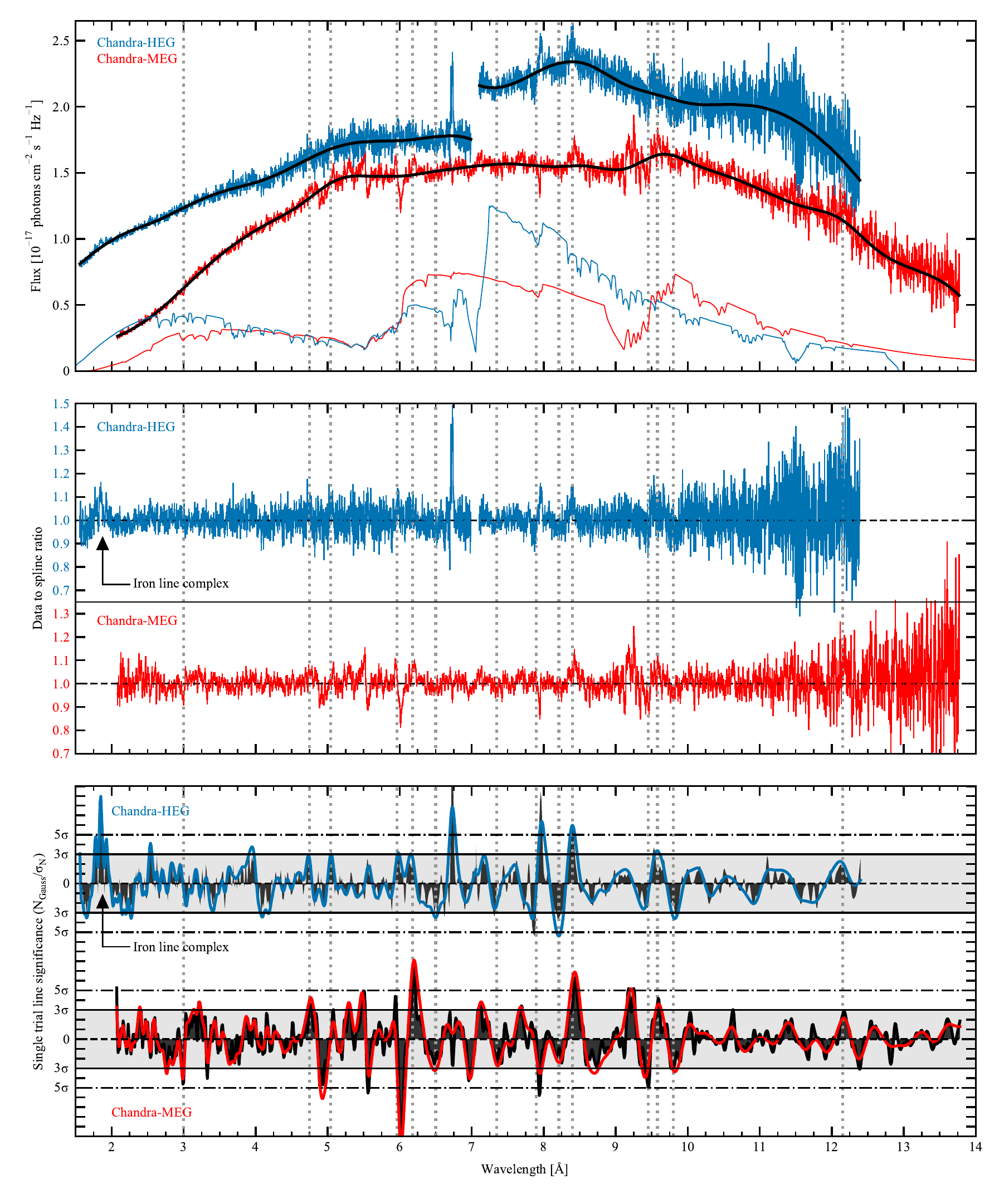}
    \caption{Overview of the line search. In all panels, the vertical dotted lines indicate possible narrow features. \textit{Top:} the HEG (blue in all panels) and MEG (red in all panels) spectra, with the spline continuum models in black. The instrument effective areas are shown in arbitrary units to indicate any response features. \textit{Middle:} the ratio of spectrum to spline model for the HEG and MEG data. The iron fluorescence complex is clearly visible below $2$ \AA. \textit{Bottom:} the single trial line significance from the line search. The solid line shows the results assuming a $2000$ km/s velocity width, while the dark grey shaded area shows the $500$ km/s search results. See also Section \ref{sec:results}.}
    \label{fig:results}
\end{figure*}

\begin{figure*}
	\includegraphics[width=\textwidth]{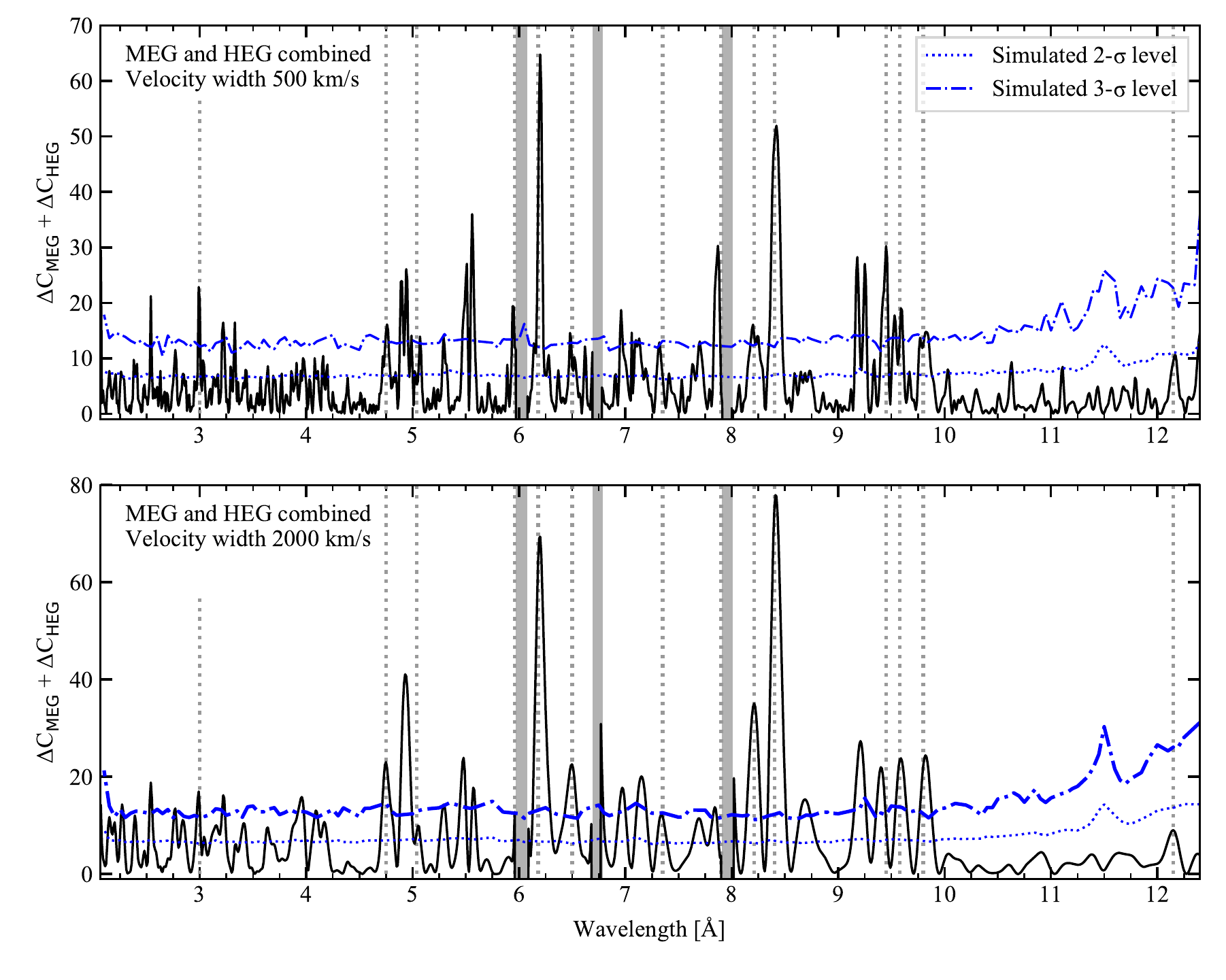}
    \caption{The summed improvement in HEG and MEG fit statistic $\Delta C$ for the $500$ km/s (top panel) and the $2000$ km/s (bottom panel) velocity width line search. The grey dotted lines show the identified emission and absorption features, while the grey bands show regions containing instrument features that are excluded to improve the clarity of the figure. Finally, the dashed and dash-dotted blue lines show the simulated $2$ and $3\sigma$ confidence levels. For the uncombined fit statistic improvement of the individual detectors, see Appendix \ref{appendix:B}}
    \label{fig:dC_comb}
\end{figure*}

\subsection{Line search results}

In Figure \ref{fig:results}, we show an overview of the analysis and the results of the line search. In the top panel, we show the HEG and MEG spectra. The large jump in the HEG spectrum around $7$ \AA~signals the complications in measuring the continuum shape discussed extensively in the previous section. The solid, black lines show the spline continuum models used in the line search. To deal with the $7$ \AA~jump, two different splines are used for the HEG spectrum, while the range between $7$ and $7.1$ \AA~is removed from the analysis. The top panel also shows the effective area shape for both detectors in arbitrary units, to indicate the instrument response. This can be used to test whether any narrow features identified in the line search might be instrumental, and shows that the $7$ \AA~jump is directly on top of the strongest HEG detector feature. We note that using the \textsc{gain} command in \textsc{xspec} to investigate the response feature did not provide a simple gain shift solution to reduce the large response residuals at for instance $\sim 7$ \AA~and $\sim 6.75$ \AA.

The middle panel shows the ratio between the spectra and the spline continuum models, providing a visual aid in searching for and confirming the physical nature of any narrow features. The results of the line search are shown in the bottom panel: we plot the single-trial significance ($N/\sigma_N$) of a Gaussian line of fixed width, added at the given energy. The red and black lines shows the $2000$ km/s velocity width search, while the black area shows the results for $500$ km/s. The $3\sigma$ and $5\sigma$ single-trial significance thresholds are shown to guide the eye. We re-emphasize that we analyzed the HEG and MEG spectra separately, to obtain independent search results that can be compared to distinguish physical lines from statistical fluctuations or instrumental features. Note that a negative significance signals an absorption feature. 

Finally, in all panels, the grey dotted lines indicate the narrow lines identified following the requirements set out in Section \ref{sec:linesearch}. Note that a clear Fe K complex is visible in the HEG spectrum below $2$ \AA~($\sim 6.4$--$7$ keV), which we do not indicate with grey lines for clarity of the figure. While this region is only covered in the HEG spectrum, the shape and centroid energies -- matching the Fe fluorescence lines expected in BeXRBs \citep[][see the next section]{torrejon10} -- of the three lines clearly show that this feature is real. All identified lines are listed in Tables \ref{tab:identifications_emission} (emission), \ref{tab:identifications_rest}, and \ref{tab:identifications_shifted} (both absorption).

\subsubsection{Significance simulations}

An alternative significance estimator for narrow features is the change in fit statistic after the addition of a narrow line at a certain wavelength, $\Delta C(\lambda)$. This estimator offers the options of combining the results from two independently analysed spectra, by linearly adding them as $\Delta C(\lambda) =  \Delta C_{\rm MEG}(\lambda) + \Delta C_{\rm HEG}(\lambda)$. We plot the combined $\Delta C$ values from the MEG and HEG detectors, for the $500$ and $2000$ km/s velocity width separately, in Figure \ref{fig:dC_comb}. The grey dotted lines indicate the same identified features as in Figure \ref{fig:results}. For visual clarity, we removed three wavelength ranges (shown by the grey bands) where large instrumental features in a single detector yield extreme $\Delta C$ values. In Appendix \ref{appendix:B}, we also show the uncombined $\Delta C$ results for each individual detector (Figures \ref{fig:B1} and \ref{fig:B2}).

To assess the significance of narrow features beyond the single-trial estimates shown in Figure \ref{fig:results}, we perform Monte-Carlo simulations of the continuum spline models. For each combination of detector (HEG or MEG) and velocity width ($500$ or $2000$ km/s), we use \textsc{XSPEC} to simulate $3000$ spectra based on the continuum spline model. For each simulated spectrum, we then repeat the line search, with a reduced resolution of $0.05$\AA~to optimize computational time. At each trial wavelength, we calculate the $2$ and $3\sigma$ confidence levels by calculating the $95.4^{\rm th}$ and $99.7^{\rm th}$ percentile of the simulated $\Delta C$ values, respectively. For the combined HEG and MEG results, we first add the simulated fit improvements, and calculate the above percentiles for this summed $\Delta C$. The simulated confidence levels are shown in Figures \ref{fig:dC_comb}, \ref{fig:B1}, and \ref{fig:B2}, as the blue dotted and dash-dotted lines, respectively.

The majority of lines identified in Figure \ref{fig:dC_comb} and listed in Tables \ref{tab:identifications_emission}, \ref{tab:identifications_rest}, and \ref{tab:identifications_shifted}, shown by the grey dotted lines, show up as $\geq 3\sigma$ significant in either the $500$ or $2000$ km/s line search (or in both). The exception is the $12.15$ \AA~emission line, which is only $\sim 2\sigma$ significant at $500$ km/s. Unsurprisingly, this line is also not formally significant in Figure \ref{fig:dC_comb}, which can be explained by the poor S/N in this region of the spectra. However, given the prominence of this feature in other BeXRBs (see Section \ref{sec:ident_em} and \ref{sec:discussion}), we include it in our further analysis. Finally, several other features appear significant above $3\sigma$. However, these all originate from a strong feature in only one of the two detectors and we therefore do not further analyse these (see also Appendix \ref{appendix:B}).

\subsection{Emission line analysis}
\label{sec:ident}

\subsubsection{Line identification}
\label{sec:ident_em}

\begin{table}
	\centering
	\caption{Identification of the detected emission lines in Sw J0243. See Section \ref{sec:ident_em} for details.}
	\label{tab:identifications_emission}
	\begin{tabular}{lll} 
\hline
$\lambda_{\rm obs}$ & Identification & Rest $\lambda$ \\ 
\hline \hline
1.77 \AA & Fe K$\beta$ & 1.77983 \AA \\
1.85 \AA & Fe XXV (He$\alpha$-like) & 1.85040 \AA \\
1.93 \AA & Fe K$\alpha$ & $1.93$ \AA \\
\hline
4.75 \AA & S XVI Ly$\alpha$ & 4.72915 \AA \\
5.04 \AA & S XV (He$\alpha$-like resonance) & 5.03873 \AA \\
5.96 \AA & Unknown & -- \\
6.18 \AA & Si XIV Ly$\alpha$ & 6.18223 \AA \\ 
8.4 \AA & Mg XII Ly$\alpha$ & 8.421 \AA \\
9.50 \AA & Ne X Ly$\delta$ & 9.48075 \AA \\
12.15 \AA & Ne X Ly$\alpha$ & 12.1339 \AA \\
\hline
	\end{tabular}
\end{table}

High-resolution X-ray observations of both super-Eddington X-ray binaries \citep{pinto16,koliopanos18b} and BeXRBs at lower accretion rates \citep{lapalombara16,grinberg17} typically reveal emission lines at rest . Indeed, six out of seven emission lines detected by our line-search algorithm can be straightforwardly identified as such. All emission line identifications are summarized in Table \ref{tab:identifications_emission}. The observed wavelengths $\lambda_{\rm obs}$ of four of these features are consistent with strong Ly$\alpha$ emission from Ne X ($\lambda_{\rm obs} = 12.15$ \AA), Mg XII ($\lambda_{\rm obs} = 8.4$ \AA), Si XIV ($\lambda_{\rm obs} = 6.18$ \AA), and S XVI ($\lambda_{\rm obs} = 4.75$ \AA). As shown in the above references, these ions are often observed in rest-emission in X-ray binaries accreting above the Eddington limit. 

This leaves three lines to be identified, at $5.04$ \AA, $5.96$ \AA, and $9.50$ \AA. The first wavelength, $5.04$ \AA, coincides with the resonance line of He$\alpha$-like S VI, which fits with the detection of the Ly$\alpha$ emission line of S XVI. The $9.50$ \AA~emission line fits with Ly$\delta$ emission of Ne X, which we know is present from the Ly$\alpha$ line. Finally, no emission lines appear to be present within $0.1$ \AA~of the $5.96$ \AA~feature and we do not identify this line. While a feature is visible in the HEG spectrum at this wavelength, the MEG feature is very close to a large instrumental residual associated with a large feature in the instrument response. It might therefore be a spurious detection. 

In addition to these seven emission lines present in both detectors, the iron fluorescence complex below $2$ \AA~is clearly visible in the HEG spectrum. We measured centroid wavelengths of $1.93$ \AA, $1.85$ \AA, and $1.77$ \AA, consistent with Fe K$\alpha$, He$\alpha$-like Fe XXV, and Fe K$\beta$, all at rest. Such emission lines at rest are also seen in all \textit{Chandra} gratings spectra of the ten HMXBs analysed in the overview work by \citet{torrejon10}. 

Analysing a \textit{NuSTAR} observation early on in the outburst of Sw J0243 (during the sub-Eddington phase), \citet{bahramian17b} report the presence of a broad ($\sigma = 0.3 \pm 0.1$ keV) Gaussian iron line at $6.42 \pm 0.07$ keV ($\sim 1.931$ \AA). This could either be the same feature as present in the HEG spectrum, only not resolved into the three individual lines. Alternatively, the \textit{NuSTAR} feature might be a relativistically broadened reflection feature, which transitioned into the three narrow lines we observe as the mass accretion rate increased. Finally, the observed HEG structure could arise from two absorption features on top of a broad emission feature. However, we could not find an satisfactory fit of the HEG spectrum with such a combination of emission and absorption. Combined with the accurate match between the centroid wavelengths and the expected iron line energies, we conclude that the final option is unlikely.

\subsubsection{Emission line modelling}
\label{sec:physmodel}

To further analyse the emission lines, we performed spectral fits with an emission line model added to the spline continuum. Such physical modelling can provide insights in the properties of the emitting gas, such as temperature and ionisation state. By comparing different models, the origin of the ionisation can also be constrained. We performed line modelling using two models in \textsc{xspec}: \textsc{BAPEC}, which models a velocity-broadened, shock-ionised gas, and \textsc{PHOTEMIS}, describing emission from a photo-ionised plasma. As we find no evidence for red or blueshifts in the identified emission lines, we freeze the redshift parameter in both models to zero. In both models, we assume Solar abundance ratios. We fit the MEG and HEG spectra simultaneously, both with their own spline continuum model, and keep the line model parameters tied between the spectra.

The \textsc{BAPEC} model provides the best fit for a velocity broadening of $v=1100^{+200}_{-340}$ km/s and a temperature $kT = 0.68 \pm 0.03$ keV, with an improvement in C-statistic of $\Delta C = 47.3$ for three extra free parameters (including normalisation). Visual inspection of the residuals reveals that this improvement largely arises from fitting the $12.15$ \AA~Ne X line, while no other emission lines are fitted. An issue with the \textsc{BAPEC} model is the presence of a significant pseudo-continuum of lines, which cannot be fitted to the non-physical continuum of the \textit{Chandra} observations. Therefore, this systematic effect prevents a more accurate fit of the spectra.

The \textsc{PHOTEMIS} model, however, provides a formally better description of the emission lines with a $\Delta C = 146.8$ for three additional free parameters, for an ionisation parameter $r\log\xi = 2.77 \pm 0.05$ and a turbulent velocity $v = (2.2 \pm 0.2)\times10^2$ km/s. The five strongest lines in the model are located at $12.15$ \AA, $8.4$ \AA, $6.18$ \AA, $5.04$ \AA, and $4.75$ \AA, fitting the Ne X, Mg XII, SI XIV, S XV, and S XVI features, respectively. The only detected narrow features not described in the model (see Table \ref{tab:identifications_emission}) are the unidentified $5.96$ \AA~feature, and the three iron lines below $2$ \AA, which are located outside the fitted wavelength range. While this suggests the emitting gas could be photo-ionised rather than shock-ionised, the comparison with the \textsc{BAPEC} model is complicated by the systematic pseudo-continuum issues in the latter.

\subsection{Absorption line analysis}
\subsubsection{No outflow scenario}
\label{sec:ident_ab1}

\begin{table}
	\centering
	\caption{Identification of absorption lines in Sw J0243 in the no-outflow scenario. See Section \ref{sec:ident_ab1} for details. Note: for the Fe ions, multiple transitions fall close to the observed wavelength. Therefore we do not list a single rest wavelength.}
	\label{tab:identifications_rest}
	\begin{tabular}{lllll} 
$\lambda_{\rm obs}$ & \multicolumn{2}{l}{Identification} & \multicolumn{2}{l}{Rest $\lambda$} \\ 
\hline \hline
\multicolumn{5}{c}{\textit{Both interpretations}} \\

3.00 \AA \hspace{7mm} & \multicolumn{2}{l}{Ca XX Ly$\alpha$ / Unknown} \hspace{7mm} & \multicolumn{2}{l}{3.02029 \AA~/ --} \\
8.21 \AA & \multicolumn{2}{l}{Fe XXI-XXIV} & \multicolumn{2}{l}{See caption} \\
9.45 \AA & \multicolumn{2}{l}{Fe XX-XXII} & \multicolumn{2}{l}{See caption} \\
9.80 \AA & \multicolumn{2}{l}{Fe XIX-XXII} & \multicolumn{2}{l}{See caption} \\ 
\hline \hline
\multicolumn{5}{c}{\textit{Fe interpretation}} \\
6.50 \AA & \multicolumn{2}{l}{Unknown} & \multicolumn{2}{l}{--} \\
7.35 \AA & \multicolumn{2}{l}{Fe XXII-XXIV} & \multicolumn{2}{l}{See caption} \\
7.90 \AA & \multicolumn{2}{l}{Fe XXII-XXIII} & \multicolumn{2}{l}{See caption} \\ 
\hline
\multicolumn{5}{c}{\textit{Fe+Mg interpretation}} \\
6.50 \AA & \multicolumn{2}{l}{Mg XII} & \multicolumn{2}{l}{6.4974 \AA} \\
7.35 \AA & \multicolumn{2}{l}{Mg XI} & \multicolumn{2}{l}{7.3101 \AA} \\
7.90 \AA & \multicolumn{2}{l}{Mg XI H$\beta$ w} & \multicolumn{2}{l}{7.8503 \AA} \\ 
\hline
	\end{tabular}
\end{table}

The identification of the detected absorption lines is more ambiguous than that of the emission lines, as the possible presence of blue-shifted lines greatly increases the feasible line identifications. Here, we will first focus on an identification scenario where no outflow was present and all lines are at rest. In this scenario, the absorption lines are either dominated by only Fe lines or by a combination of Fe and Mg lines.

Out of the seven detected absorption lines, listed in Table \ref{tab:identifications_rest}, four can be identified with the same ions in both the only-Fe and Fe+Mg interpretations. Iron absorption can account for the features observed at $8.21$ \AA, $9.45$ \AA, and $9.80$ \AA, where several transitions of respectively Fe XXI-XXIIV, Fe XX-XXII, and Fe XIX-XXII are located. The $3.00$ \AA~absorption line could be associated with the Ly$\alpha$ transition of Ca XX at rest. However, while iron is often invoked to explain observed narrow lines in X-ray binaries \citep[e.g][]{pinto16} and the clear iron fluorescence lines show that iron is present in Sw J0243 (c.f. Section \ref{sec:ident_em}), Ca XX is not typically observed in these systems. Additionally, the $3.00$ \AA~feature appears predominantly present in the MEG spectrum. Therefore, there is a possibility that this line is merely a spurious detection. 

Assuming no blue-shifted absorption, we can link the other three detected lines with either iron (the \textit{Fe interpretation} in Table \ref{tab:identifications_rest}) or magnesium (the \textit{Fe+Mg interpretation}). For the former interpretation, the $7.35$ \AA~and the $7.90$ \AA~lines can be associated with Fe XXII-XXIV and Fe XXII-XXIII, respectively. This would leave the $6.50$ \AA~feature unidentified. For the Fe+Mg interpretation, this $6.50$ \AA~feature could arise from Mg XII, while the $7.35$ \AA~and the $7.90$ \AA~lines would be Mg XI and the H$\beta$-like resonance of Mg XI, respectively. However, this Fe+Mg interpretation has several caveats: while the wavelengths match up and the $8.4$ \AA~Mg XII Ly$\alpha$ emission line shows that Mg is present, it is unexpected that Mg XII would be observed at rest in both emission and absorption \citep{grinberg17}. Furthermore, the $6.50$ \AA~Mg XII and $7.35$ \AA~Mg XI lines are relatively weak transitions, and one would therefore expect to see other or a larger number of Mg absorption lines instead. Finally, the rest and observed wavelengths do not match up perfectly in this interpretation. 

For the above two interpretations, we can calculate the cumulative improvement in fit statistic by adding the combined $\Delta C$ values for each identified line. The six identified lines in the Fe interpretation yield a cumulative $\Delta C$ of $127.1$ ($124.5$) for 500 (2000) km/s, while the seven identified lines in the Fe+Mg interpretation sum up to $\Delta C = 141.7$ ($147.0$) for 500 (2000) km/s.

\subsubsection{Outflow scenario}
\label{sec:ident_ab2}

\begin{table}
	\centering
	\caption{Identification of a selection of absorption lines in Sw J0243 in the outflow scenario. See Section \ref{sec:ident_ab2} for details. The remaining absorption lines are interpreted as in Table \ref{sec:ident_ab1} and \ref{sec:ident_ab1}.}
	\label{tab:identifications_shifted}
	\begin{tabular}{llll} 
\hline
Observed $\lambda$ & Ion & Rest $\lambda$ & Shift \\ 
\hline \hline
6.50 \AA & Mg XII Ly$\alpha$ & 8.421 \AA & $-0.226c$ \\
7.35 \AA & Ne X Ly$\delta$ & 9.48075 \AA & $-0.225c$ \\
9.45 \AA & Ne X Ly$\alpha$ & 12.1339 & $-0.222c$\\
\hline
	\end{tabular}
\end{table}

Alternatively, we consider a scenario where a selection of the absorption features are identified as blue-shifted transitions with the same outflow velocity. The firm detection of Ne, Mg, Si and S in emission aids in this approach, as it provides a starting point to identify lines that might be expected in absorption. In fact, in the discovery of ultra-fast outflows (UFOs) in ULXs, \citet{pinto16} observed many of the observed rest emission features in absorption with the same blue-shift. For instance, in these ULXs, the Ne X Ly$\alpha$ rest emission line that is also present in Sw J0243, is also observed in absorption with a $\sim -0.2c$ velocity shift.

To test for a similar scenario in Sw J0243, we calculated the blue-shifts required to explain every combination of an observed absorption line and an observed higher-wavelength emission line (excluding the unidentified $5.96$ \AA~emission feature). In the case of an outflow, we would expect a similar blue-shift to appear for a number of such pairings. Indeed, for an outflow velocity of $\sim 0.22c$, the $6.50$~\AA, $7.35$~\AA~and $9.45$~\AA~absorption lines can be linked to, respectively, the observed $8.4$~\AA~Mg XII Ly$\alpha$ line, $9.50$~\AA~Ne X Ly$\delta$ line, and $12.15$~\AA~Ne X Ly$\alpha$ line (see Table \ref{tab:identifications_shifted}). This scenario provides a seemingly more feasible explanation of the $6.50$~\AA~feature than that in Section \ref{sec:ident_ab1}, while the required velocity is similar to that of UFOs in ULXs with an unknown accretor \citep{pinto16} and the ULX pulsar NGC 300 ULX-1 \citep{kosec18b}. The cumulative $\Delta C$ for these three blueshifted lines is $57.7$ ($56.8$) for a velocity width of 500 (2000) km/s.

If Sw J0243 launches an outflow with a velocity of $0.22$c, we can ask two more questions. Firstly, why do we only observe the Ne X and Mg XII emission lines in absorption as well? Shifting the two S XVI and the Si XIV emission lines by the same velocity returns wavelengths of $5.45$ \AA, $4.15$ \AA, and $4.44$ \AA. Out of these, hints for an absorption feature can only be seen around $\sim 4.15$ \AA~in the HEG spectrum. However, this is not a convincing feature, and no hints of a line are present at the other two wavelengths. Given the strength of the Ne X and Mg XII lines in HMXBs in general, and in Sw J0243 specifically, it is however not surprising that these ions are most easily detected in blue-shifted absorption. 

Secondly, we consider whether any of the other absorption features might be associated with $0.22c$ blue-shifted lines from species not observed in emission. Shifting these four remaining absorption lines, only the $8.21$ \AA~feature yields a possible match; its shifted wavelength of $9.33$~\AA~is similar to the $9.31$ \AA~forbidden transition of He$\alpha$-like Mg XI. However, it appears unlikely that only this forbidden line is observed, while other stronger transitions are not seen. Therefore, the only direct evidence for the outflow consists of the three absorption lines listed in \ref{tab:identifications_shifted}, and we interpret the remaining absorption features as in Section \ref{sec:ident_ab1}.

\section{Discussion and conclusions}
\label{sec:discussion}

We have reported high-resolution \textit{Chandra} X-ray spectroscopy of the super-Eddington outburst of Sw J0243. A search for narrow emission and absorption features reveals a number of both, present in both the HEG and MEG spectrum. The emission lines can be identified with Fe, S, Si, Mg, and Ne ions at rest. The absorption features can either be interpreted to be all at rest (from Fe and possibly Ca and Mg), or a combination of some lines at rest and three blue-shifted Mg and Ne absorption lines at $v \approx -0.22c$. Here, we briefly review our method, discuss the possible outflow in the context of ULXs and close-by BeXRBs, and finally present future improvements for the study of outflows from BeXBRs during super-Eddington phases.  

\subsection{Line-search method}

As shown in Figure \ref{fig:crappycontinuum}, the continuum of the \textit{Chandra} spectrum differs greatly between detectors and deviates from the shape measured by \textit{Swift}. However, the iron fluorescence complex below $2$~\AA, i.e. around $6.5$ keV that is observed in all HMXBs with \textit{Chandra} gratings observations \citep{torrejon10}, is clearly detected. Similarly, the Ne X and Mg XII Ly$\alpha$ emission lines at respectively $12.15$~\AA~and $8.4$~\AA, often observed in neutron star HMXBs \citep[e.g.][]{grinberg17,koliopanos18b,lapalombara16}, are visible in the spectra even by eye (c.f. the top panel of Figure \ref{fig:results}). This suggests that indeed, while the continuum is affected by the observing setup, narrow features remain detectable \citep{schulz09}. In addition, the consistency checks of the spline continuum model (e.g. Appendix \ref{sec:app_checks}) show that none of the detected lines or conclusions are due to the non-physical nature of this model. 

Our adopted line-search method follows the rationale first used by \citet{pinto16}, and later applied to both \textit{XMM-Newton} RGS and \textit{Chandra} observations by \citet{degenaar17}, \citet{kosec18b}, \citet{kosec18a}, and \citet{vandeneijnden18c}. However, as discussed in more detail in \citet{vandeneijnden18c}, estimating formal significances of detected features is challenging. The significances shown in Figure \ref{fig:results} are single-trial values, while estimating the number of independent trials is difficult: fitting a Gaussian line at neighbouring wavelength gridpoints is not independent, as the Gaussian width exceeds the resolution of the grid. For a velocity width of $500$ km/s, a Gaussian line covers between 4 and 28 wavelength bins of $0.01$~\AA~width in its one-sigma range, where the number varies since a constant velocity width translates to a variable width in wavelength-space. For the $2000$ km/s search, these numbers are multiplied by four. Therefore, while we fit the normalization of a Gaussian line at 1161 wavelength bins, a much smaller -- but difficult to estimate precisely -- number of those trials are truly independent.

Therefore, we also performed Monte-Carlo simulations of the continuum shape to estimate how likely random fluctuations can reproduce the observed excesses \citep[e.g][]{vandeneijnden17,kosec18b}. Secondly, we opted for an independent analysis and subsequent comparison of the HEG and MEG spectrum, since statistical fluctuations or response-effects are less likely to show up in both detectors at the same wavelength. Thirdly, searching with two different velocity widths decreases the probability of statistical fluctuations being identified as a line: while a small number of bins fluctuating either above or below the continuum by chance might mimick a narrow line, it would not be identified as such when searching with a broader velocity width. Finally, it is important that any possible line can be identified within a coherent physical picture of the system. For this reason, we do not identify for example the possible line at $5.96$ \AA.

\subsection{An ultra-fast outflow from Sw J0243?}

While the combination of detected absorption lines can be interpreted as simply Fe (and Mg) ions at rest, a more interesting possibility is the presence of an outflow suggested by the presence of absorption features at a $\sim 0.22c$ blue-shift from Mg Ly$\alpha$, Ne X Ly$\alpha$, and Ne X Ly$\delta$ -- which are all observed in emission. The presence of such an outflow during the super-Eddington regime fits both theoretical predictions and simulations \citep[e.g.][]{shakura73,ohsuga11,mckinney14,mckinney15,hashizume15,kingmuldrew16} and observational work \citep[e.g.][]{lee02,pinto16,kosec18b,allen18}. 

In the outflow scenario, the blueshifted absorption lines originate from the same ions as several rest-emission lines. This combination of rest emission and blue-shifted absorption from the same ions might arise from the outflow, seen in absorption, shocking with the surrounding medium, seen in emission. Such a scenario is also invoked in \citet{pinto16}, where the emission can be modeled as a shock-ionised gas. Alternatively, the emission might arise from a different region in the system, instead of the outflow but with the same ions present \citep{grinberg17}, such as the accretion flow.

The Sw J0243 high-resolution X-ray spectrum is similar to that of other Galactic and SMC BeXRBs in several aspects. For instance, the \textit{XMM-Newton} RGS spectrum of SMC X-3 during its super-Eddington state also shows rest-emission lines of Ne X, Mg XII, and Fe XXIII-XXIV \citep{koliopanos18b}. Additionally, a possible blue-shifted Mg XII absorption line is detected, which would imply an outflow with a $\sim 0.07c$ velocity. The presence of Mg XII in both rest emission and blue-shifted absorption mirrors our outflow interpretation for Sw J0243. At slightly sub-Eddington X-ray luminosity ($L_X \sim 10^{38}$ erg/s), the BeXRB SMC X-2 shows both rest emission lines of Ne X and Si XIV \citep{lapalombara16}, as we also identify in Sw J0243 (in addition to several other rest emission lines not seen in Sw J0243 due to the difference between the HEG/MEG and RGS bandpasses). However, no hints for absorbtion features or an outflow are present. Even further down in the sub-Eddington regime, \citet{grinberg17} report a plethora of rest emission lines in Vela X-1, including the Fe, Si, Mg, and Ne species identified in Sw J0243. But again, no outflow is detected, despite the high-quality observations which would likely reveal an outflow similar to the one we possibly detect in Sw J0243. 

Given the super-Eddington accretion rate of Sw J0243 during the \textit{Chandra} observation, ULXs form a second interesting source class for comparison. \citet{kosec18b} reported the detection of a possible outflow from the ULX pulsar NGC 300 ULX-1. This outflow was observed through the identification of blueshifted O XVII and O XVIII absorption lines in RGS spectra, which fall outside the \textit{Chandra} bandpass. The inferred velocity of $0.22c$ is however consistent with the velocity of the possible wind in Sw J0243. In addition, \citet{pinto16} present the discovery of $\sim 0.2c$ outflows from the unclassified ULXs NGC 1313 X-1 and NGC 5408 X-1; interestingly, next to the similarity of the wind velocity, both sources show a combination of emission lines at rest, absorption lines from the same species at a blue-shift, and additional rest absorption lines. This mimicks exactly our outflow scenario for Sw J0243. Finally, an outflow with a higher velocity of $\sim 0.34c$ was tentatively claimed in NGC 5204 X-1 by \citet{kosec18a}, but this result awaits confirmation. 

Around the time of the \textit{Chandra} observation of Sw J0243, the source also launched a radio jet \citep{vandeneijnden18d}. Radio observations taken four days later show an optically-thin radio spectrum, implying that during this super-Eddington state, the jet consisted of discrete ejecta \citep[e.g.][]{fender06}\footnote{We note that, as discussed extensively in \citet{vandeneijnden18d} and \citet{vandeneijnden19}, the observed radio properties (flux densities, spectral shape, and evolution) throughout the entire outburst show that this radio emission can not originate from either a stellar or disk wind.}. While there is no simultaneous radio and high-resolution X-ray coverage, Sw J0243 remained in a very similar state between the jet and possible wind detection. Therefore, we deem it likely that both an ultra-fast outflow and a jet are launched at the same time. Such behaviour is observed more often during the super-Eddington regime in other sources: black holes and Z-sources also show winds and jets during the same super-Eddington accretion states \citep{homan07,homan16,allen18}. However, while both winds and jets have been inferred in ULXs \citep[e.g.][]{middleton13,cseh14,kaaret17}, these have never been observed in the same target, let alone at the same time. Given the difficulty to find these outflows \citep[e.g.][]{kosec18a}, due to the large distances to ULXs, Galactic BeXRBs offer a new avenue to explore the connection between winds and jets in the super-Eddington regime. 

One particularly puzzling aspect of the jet launched by Sw J0243 is its faintness compared to fast-spinning, weakly-magnetized accreting neutron stars at similar super-Eddington accretion rates \citep{vandeneijnden18d}. This can naively be explained by the slow spin of Sw J0243 \citep[e.g.][]{parfrey16}; however, at lower accretion rates, the radio brightness of Sw J0243 is consistent with faster-spinning neutron stars, which implies a more complicated picture \citep{vandeneijnden19}. Possibly, the presence of an ultra-fast disk wind during the super-Eddington phase of the outburst can regulate the jet power; depending on its launch radius, a wind might decrease the mass accretion rate in the inner accretion flow and reduce the matter available to form the jet. Alternatively, it might carry away excess angular momentum and reduce the jet power. A similar interplay between the wind and the jet has earlier been proposed to explain the jet-wind dichotomy in GRS 1915+105 (\citealt{neilsen09}, see also \citealt{diaztrigo13}). Since the jet in Sw J0243 is consistent with the population of other NS jets at lower X-ray luminosity, this scenario assumes that the ultra-fast outflow was driven by the super-Eddington accretion rate and disappeared as the outburst decayed. 

A scenario where the presence of a strong wind outflow regulates the jet power does not occur in Z-sources: these sources launch powerful jets with the highest radio brightness of any type of accreting neutron star. However, while the winds in these systems can carry away significant amounts of mass \citep{ponti12,allen18}, they generally do not reach velocities similar to those in ULXs and inferred here for Sw J0243 \citep[i.e. maximally one per cent of the speed of light;][]{diaztrigo13}. In addition, the inner accretion flow and jet launching regions differ greatly between weakly-magnetized Z-sources and the more strongly-magnetized ULX pulsars and BeXRBs \citep[e.g][]{mushtukov17,walton18a}. Finally, while Z-sources accrete close to or above the Eddington limit, our Sw J0243 \textit{Chandra} observation was taken at one order of magnitude higher X-ray luminosity. Therefore, any coupling between the (super)-Eddington winds and jets would not necessarily be the same in Sw J0243, ULX pulsars, and Z-sources.

\subsection{Future observations}

While the presence of an ultra-fast outflow from Sw J0243 fits with the \textit{Chandra} high-resolution X-ray spectrum, the continuum issues complicate a full analysis and more detailed physical modelling than described in Section \ref{sec:physmodel}. The possible wind detection does however showcase the power of studying Galactic BeXRBs for understanding pulsating ULXs and their outflows. Therefore, future observational campaigns combining radio, X-ray and UV observations would be highly valuable: dense radio monitoring can track the jets, while high-resolution X-ray and UV spectra (from i.e. the \textit{Hubble Space Telescope}) taken at different phases in the outburst can track the onset and evolution of any wind outflow. Through such detailed monitoring, the relation between the jet and wind can be studied as well, for instance aiming to understand if and how the (presence of the) wind might influence the jet power. These future observations can also reveal how commonly super-Eddington BeXRBs launch a wind and jet simultaneously, to better understand the expected outflow properties of ULX pulsars. 

\section*{Acknowledgements}

We are grateful to the anonymous referee for their constructive comments that improved this paper, and to Belinda Wilkes and the \textit{Chandra} scheduling team for rapidly accepting and performing the Director's Discretionary Time observation. We thank Ciro Pinto, Mark Reynolds, and Jacco Vink for useful discussions on X-ray spectroscopy, super-Eddington accretion, and outflows. JvdE, ND, and JVHS are supported by a Vidi grant from the Netherlands Organization for Scientific Research (NWO) awarded to ND. TDR is supported by a Veni grant from the NWO. COH acknowledges an NSERC Discovery Grant. This work made use of data supplied by the UK Swift Science Data Centre at the University of Leicester. This research has made use of MAXI data provided by RIKEN, JAXA and the MAXI team. ASAS-SN is supported by NSF grant AST-1515927. Development of ASAS-SN has been supported by NSF grant AST-0908816, the Center for Cosmology and AstroParticle Physics at the Ohio State University, the Mt. Cuba Astronomical Foundation, and by George Skestos.




\input{references.bbl}



\appendix

\section{Checks of the spline continuum}
\label{sec:app_checks}

\begin{figure*}
	\includegraphics[width=\textwidth]{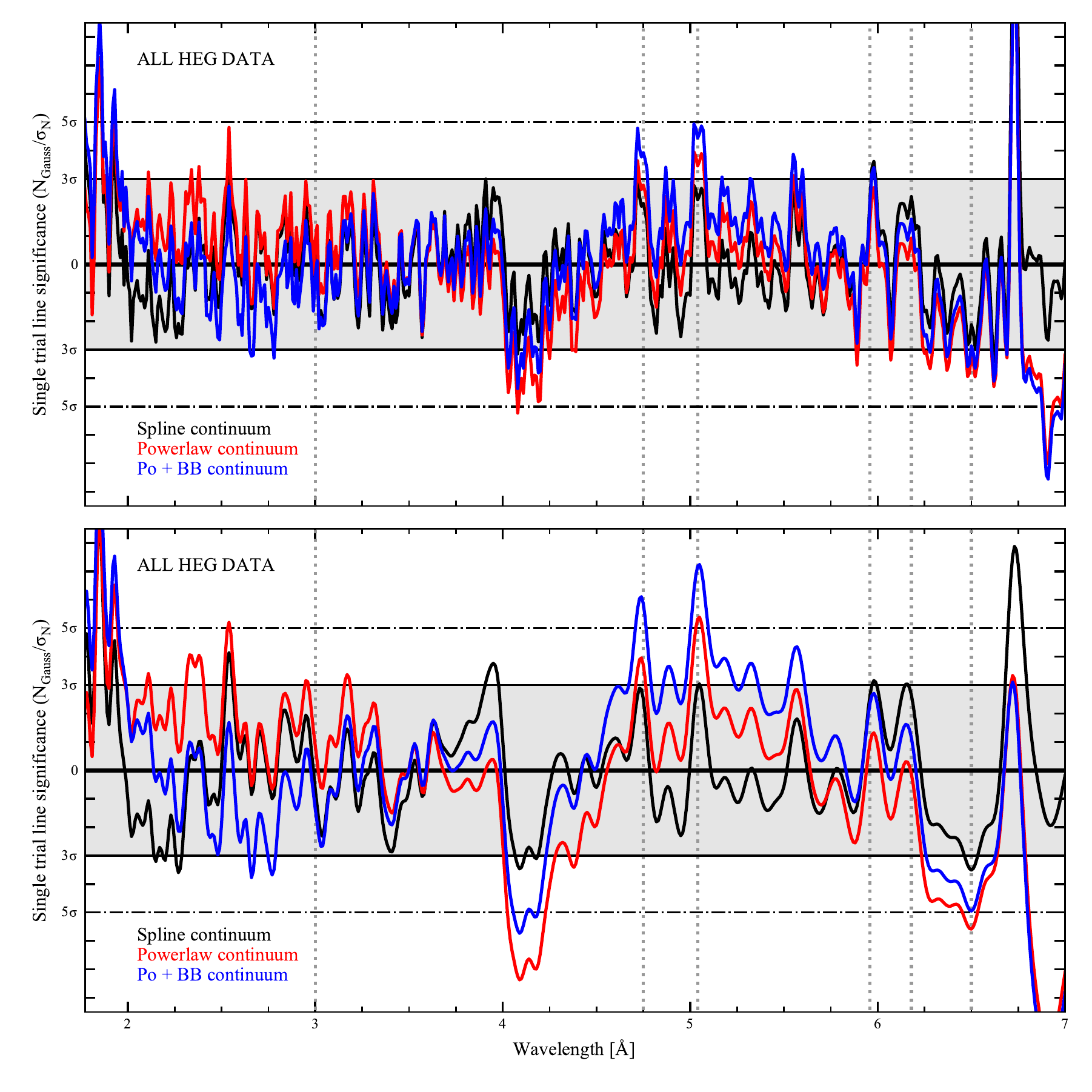}
    \caption{The first consistency check of the spline continuum, shown the HEG spectrum below $7$ \AA: both panels show the results of the line search algorithm for different velocity widths (top: $500$ km/s, bottom: $2000$ km/s). The black curves use the spline continuum model, while the red and blue curves use a power law and a power law + blackbody continuum model, respectively. In both cases, and especially in the bottom panel, residual trends remain when using the physical continuum, which can artificially enhance single-trial significances. However, the individual narrow features appear for all continuum models.}
    \label{fig:check_phys}
\end{figure*}

\begin{figure*}
	\includegraphics[width=\textwidth]{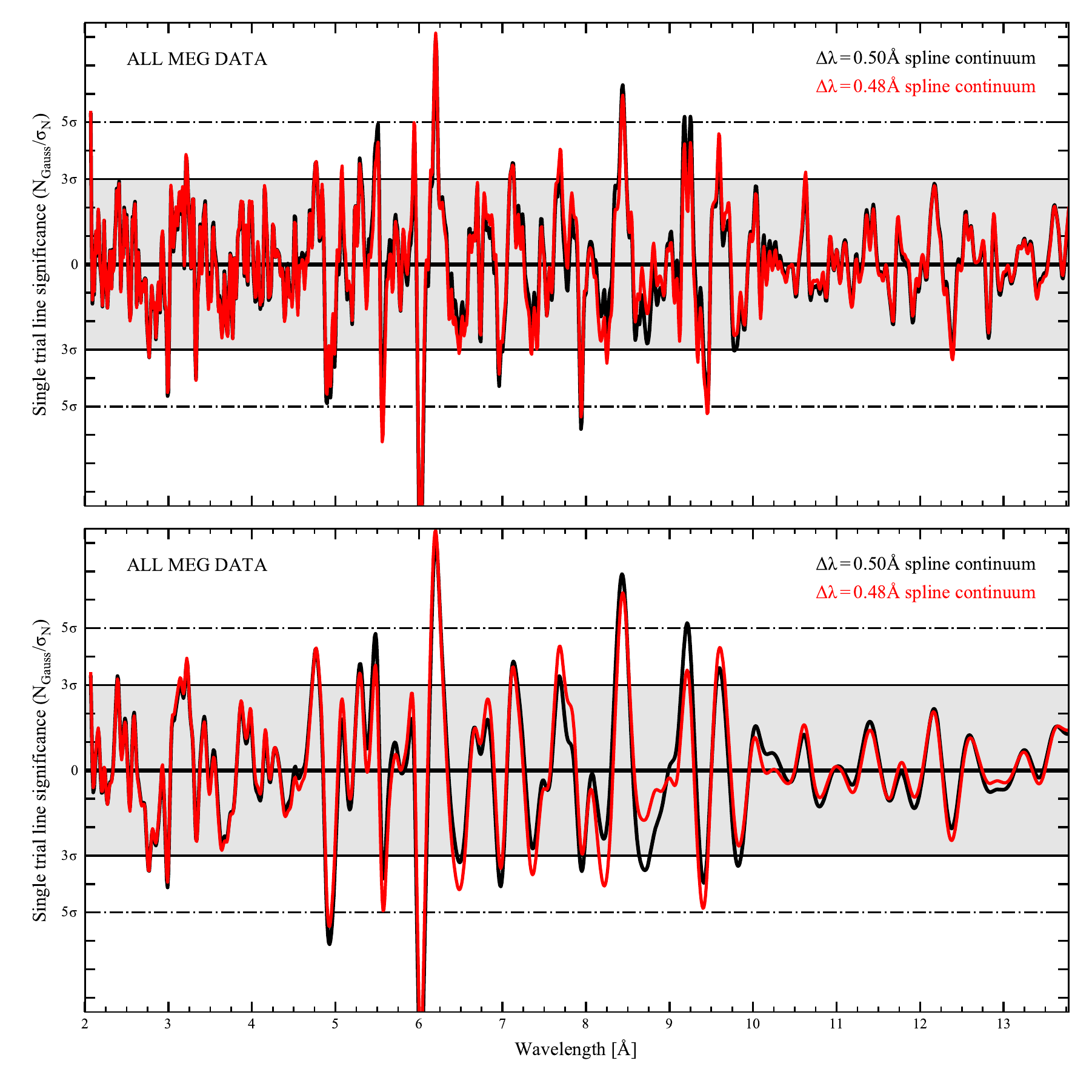}
    \caption{The second consistency check of the spline continuum, shown for the MEG spectrum: both panels show the results of the line search algorithm for different velocity widths (top: $500$ km/s, bottom: $2000$ km/s). The black curves use the spline continuum model with a $0.50$\AA~ step size, while the red curve uses a slightly smaller $0.48$\AA~step size -- interpolating between different spectral bins. For both velocity widths, the results are largely consistent, with only slight differences between the two continuum models. None of the possible narrow features are affected by these small deviations.}
    \label{fig:check_shift}
\end{figure*}

Since comparisons between the HEG and MEG detectors, and with \textit{Swift} spectra taken at similar times, reveals that the \textit{Chandra} continuum is not accurate, we have performed careful checks of the validity of our spline continuum model. In Figures \ref{fig:check_phys} and \ref{fig:check_shift}, we show these checks visually. In both figures, we show the results of the line search method, similar to the bottom panel of Figure \ref{fig:results}. As described in the main text, we compare our line search results (the black lines in both panels of both figures) with the results from using different continuum models: two physical continuum models fitted jointly to the quasi-simultaneous \textit{Swift} spectra (Figure \ref{fig:check_phys}) and a spline continuum with slightly smaller step size (Figure \ref{fig:check_shift}). 

The used spline continuum appears robust during both tests: using a more physical continuum finds the same narrow features, but contains residual trends in the significance as function of wavelengths (Figure \ref{fig:check_phys}). These trends artificially enhance any narrow line significances, and signal that slight difference between the \textit{Swift} and HEG spectrum remain even between $1.75$ and $7$ \AA. When we use a smaller stepsize (Figure \ref{fig:check_shift}), we find results consistent with our original line search. This implies that these results are not affected by the possibility that the spline connects statistical outliers or narrow line features instead of the continuum. 

\section{$\Delta C$ results from individual detectors}
\label{appendix:B}

\begin{figure*}
	\includegraphics[width=\textwidth]{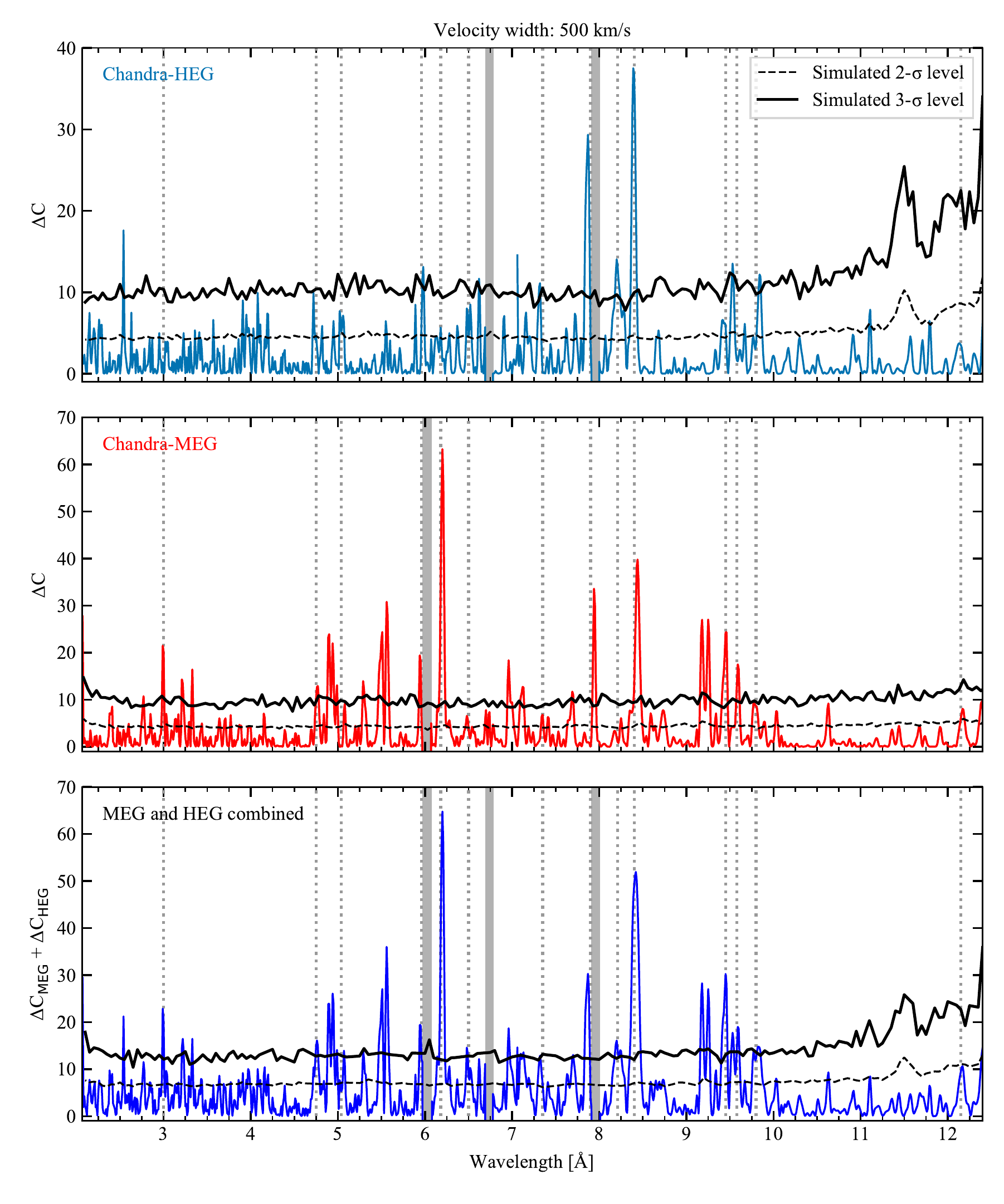}
    \caption{The improvement in fit statistic $\Delta C$ for the $500$ km/s line width search of the HEG (top panel) and MEG (top panel). The bottom panel shows the combined improvement in both detectors, as is also shown in Figure \ref{fig:dC_comb}. The grey dotted lines show the identified emission and absorption features, while the grey bands show regions containing instrument features that are excluded to improve the clarity of the figure. Finally, the solid and dashed black lines show the simulated $2$ and $3\sigma$ confidence levels.}
    \label{fig:B1}
\end{figure*}

\begin{figure*}
	\includegraphics[width=\textwidth]{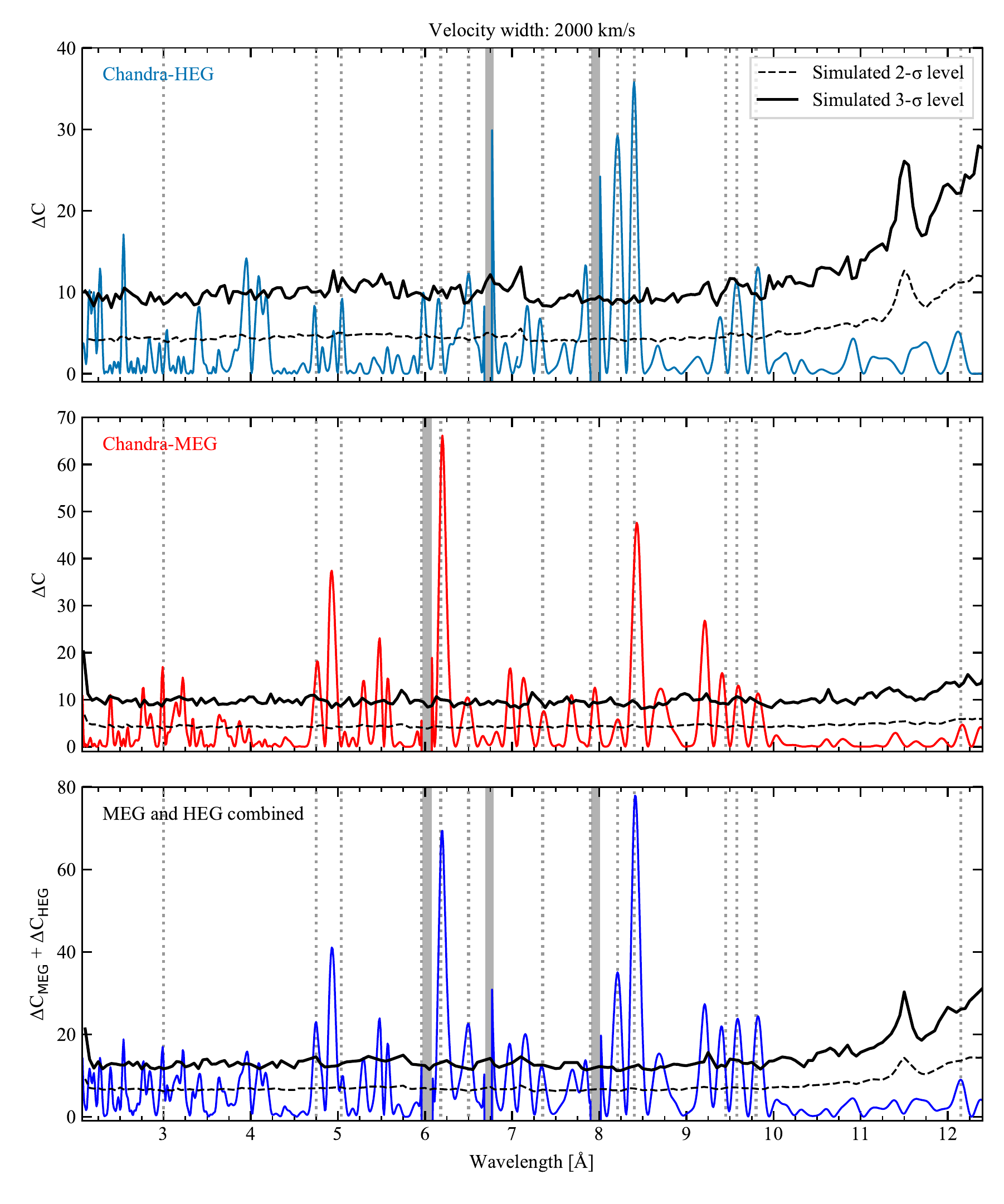}
    \caption{Same as Figure \ref{fig:B1} for the $2000$ km/s line width search.}
    \label{fig:B2}
\end{figure*}

In Figures \ref{fig:B1} and \ref{fig:B2}, we show the $\Delta C$ search results for the HEG and MEG detectors separately (upper and middle panel) and combined (lower panel). The simulated $2$ and $3\sigma$ confidence levels are shown as the black dashed and solid lines, respectively. For details on these simulations, see Section \ref{sec:results} in the main paper. Figure \ref{fig:B1} shows the results for the $500$ km/s velocity width, while Figure \ref{fig:B2} shows the $2000$ km/s results. The comparison of all three panels shows how several of the apparently significant features in the combined (bottom) panel, arise from strong (instrumental) features in only one detector -- see for instance the double-peaked feature around $9.20$ \AA~in the bottom panel of both figures, that is only present in the MEG data.

\bsp	
\label{lastpage}
\end{document}